\documentstyle[preprint,tighten,aps]{revtex}    
\begin{document}           %
\draft
\preprint{\vbox{\noindent
To appear in Physics Letters B\hfill INFNCA-TH-94-1\\
          \null\hfill MIT-CTP\#2298\\
          \null\hfill hep-ph/9404275}}
\title{Exact Sum Rules at Finite Temperature and Chemical Potential\\
       and Their Application to QCD}

\author{Suzhou Huang$^{(1,2)}$\cite{email}
        and Marcello Lissia$^{(1,3)}$\cite{email} }
\address{
$^{(1)}$Center for Theoretical Physics, Laboratory for Nuclear Science
and Department of Physics, \\
Massachusetts Institute of Technology, Cambridge, Massachusetts 02139\\
$^{(2)}$Department of Physics, FM-15, University of Washington,
Seattle, Washington 98195~\cite{present}\\
$^{(3)}$Istituto Nazionale di Fisica Nucleare,
via Ada Negri 18, I-09127 Cagliari, Italy~\cite{present}\\
and Dipartimento di Fisica dell'Universit\`a di Cagliari,
I-09124 Cagliari, Italy
         }
\date{December 1994 (revised version, original April 1994)}
\maketitle                 
\begin{abstract}
Within the framework of the operator product expansion (OPE) and
the renormalization group equation (RGE), we show that the
temperature and chemical potential dependence of the zeroth moment
of a spectral function (SF) is completely
determined by the one-loop structure of an asymptotically free theory.
This exact result constrains the shape of SF's,
and implies a highly non-trivial functional form for the SF near
second order, or weak first order, phase transitions.
Phenomenological parameterizations of the SF, often used
in applications such as the analysis of lattice QCD data or QCD
sum rule calculations at finite temperature and baryon density, must
satisfy these constraints.
\end{abstract}
\pacs{Keywords: Sum rules, Finite temperature field theory, QCD}

\narrowtext
\maketitle
Hadronic properties at finite temperature and baryon density are
of great importance in the phenomenology of heavy ions collisions,
star interior and the early universe. Moreover, the theoretical
expectation of transitions to a chirally symmetric phase and,
perhaps, to a quark-gluon plasma phase contributes to the interest
in studying the effect of matter and temperature on the QCD vacuum.

  Our present understanding of QCD at finite temperature ($T$) and
baryon density (or chemical potential $\mu$) is mainly limited
in the Euclidean realm, due to the lack of non-perturbative and
systematic calculating tools directly in the Minkowski space.
Typical methods, with QCD Lagrangian as the starting point, are
the OPE and lattice simulations. Because these two formulations
are intrinsically Euclidean, only static quantities are conveniently
studied. In order to gain dynamical informations, which are more
accessible experimentally, the analytic structure implemented
through dispersion relations often have to be invoked within the
theory of linear response.

The real-time linear response to an external source
coupled to a renormalized current $J(x)$ is given by the retarded
correlator:
\begin{equation}
K(x;T,\mu)\equiv
\theta(x_0)\langle [J(x),J(0)]\rangle_{T,\mu}\, ,
\label{corr}
\end{equation}
where the average is on the grand canonical ensemble with
temperature $T$ and chemical potential $\mu$. Disregarding possible
subtraction terms, we can write the following dispersion relation
for the frequency dependence of the retarded correlation function:
\begin{equation}
\tilde{K}(\omega,\bbox{k};T,\mu)=\int_{0}^\infty du^2 \,
{\rho(u,\bbox{k};T,\mu)\over u^2-(\omega+i\epsilon)^2} \, .
\label{disp}
\end{equation}
For convenience, we discuss only the uniform limit (${\bf k}=0$).
Upon analytic continuation, $\omega\rightarrow iQ$, the dispersion
relation in principle connects the correlator in Euclidean region
to the SF, which embodies all the real-time information.

  Excluding special cases,  where the SF
$\rho(u;T,\mu)$ is experimentally measured, neither side
of Eq.(\ref{disp}) is completely known. In order to extract
physical information in terms of elementary
excitations, the standard approach consists of calculating the
left-hand side of Eq.(\ref{disp}) (typical means are current algebras,
lattice simulations and the OPE), assuming a phenomenological motivated
fit to the SF $\rho(u;T,\mu)$ and using the dispersion relation,
Eq.(\ref{disp}), to determine the  physically motivated parameters of
the fit. This approach has been very successful at zero $(T,\mu)$,
due to the fact that we have a good qualitative understanding on
how to parameterize SF's. Two important
example are the QCD sum rules pioneered by SVZ~\cite{SVZ,QCDSR_rev},
and the analysis of lattice QCD data~\cite{chu93}.

The QCD sum rules approach has been extended also to systems at finite
temperature~\cite{Bochkarev86,Dosch88,hatsuda}.
The lack of experimental data, and of reliable nonperturbative calculations
has prompted people to use the same kinds of parameterizations that have
worked so well at zero $(T,\mu)$ with, at most, perturbative corrections.
We believe that physical results at finite $(T,\mu)$ can be strongly
biased by this assumption.
In fact, recent interpretations of lattice simulation
data~\cite{hashimoto,karsch,chu94} appear to indicate the existence
of such problems.

The purpose of this letter is to derive exact sum rules that
constrain the variation of SF's with $T$ and $\mu$.
The derivation, based on the OPE and the
RGE, has closer analogies to the analysis of deep inelastic lepton
scattering experiments than to the QCD sum rule
approaches~\cite{SVZ,QCDSR_rev,Bochkarev86,Dosch88,hatsuda}.
Moreover, our derivation is not based on expansion for large or small
$(T,\mu)$, making these sum rules valid everywhere in the $(T,\mu)$-plane,
and in particular near phase transitions.
In addition, we apply these sum rules to the chiral phase transition,
and demonstrate that SF's in some channels are
drastically modified compared both to their zero $T$ and
perturbative shape, at least when the phase transition is second order or
weak first order, confirming our worries about the non-trivial
effect of finite $T$ or baryon density.

In an asymptotically free theory, the OPE yields, {\em e.g.} in
the $\overline{\rm MS}$ scheme, the large-$Q^2$ asymptotic expansion
\begin{equation}
\tilde{K}(iQ;T,\mu,\kappa) \sim \tilde{K}_0(iQ,\kappa)+
\sum_n C_n(Q^2,g^2(\kappa),\kappa)
\langle [O_n]_\kappa\rangle_{T,\mu} \, ,
\label{ope}
\end{equation}
where $g^2(\kappa)$, $[O_n]_\kappa$'s and $C_n$'s are, respectively,
the coupling constant, the renormalized composite operators and their
corresponding Wilson coefficients at the subtraction mass scale
$\kappa$. It is important to notice that the information of the
ensemble average is encoded in the matrix elements of the composite
operators, while the Wilson coefficients and $\tilde{K}_0$ are
independent of $T$ and $\mu$. Although the matrix elements
$\langle [O_n]_\kappa\rangle_{T,\mu}$ cannot be determined perturbatively,
the $Q^2$-dependence of Wilson coefficients $C_n$ is dictated by the
renormalization group equation~\cite{brown}, and given by
\begin{equation}
C_n(Q^2,g(\kappa),\kappa)=
{c_n\bigl(g^2(Q)\bigr)\over Q^{d_n}}\,
\biggl[{g^2(Q)\over g^2(\kappa)}
\biggr]^{(2\gamma_J-\gamma_n)/2b}
\Bigl\{1+{\cal O}(g^2(Q))\Bigr\}\, ,
\label{cn}
\end{equation}
where $d_n$ is the canonical dimension of the operator $O_n$ minus
the dimension of $\tilde{K}$, and $c_n(g^2(Q))$ is calculable
perturbatively. The pure numbers $\gamma_i$ ($i=J$, $n$) and $b$
are related to the anomalous dimensions of $J$, $O_n$ and to the
$\beta$-function as follows
\begin{equation}
\Gamma_i=-\gamma_i g^2+{\cal O}(g^4)\, ,\,\,\, \text{and}
\,\,\, \beta=-b g^4+{\cal O}(g^6) \, .
\label{rgef}
\end{equation}

  To study the dependence of $\tilde{K}$ on $(T,\mu)$ we only need
to consider the difference
$
\Delta\tilde{K}(iQ) \equiv \tilde{K}(iQ;T,\mu)-\tilde{K}(iQ;T',\mu')
$
and
\begin{equation}
\Delta\tilde{K}(iQ)
=\int_0^\infty du^2 {\Delta\rho(u)\over u^2+Q^2}\, ,
\label{ddisp}
\end{equation}
where $\Delta\rho(u)\equiv\rho(u;T,\mu)-\rho(u;T',\mu')$.
This subtraction is crucial to remove $\tilde{K}_0(iQ,\kappa)$,
which contains all the terms not suppressed by a power of $1/Q^2$,
and also to make $\Delta\tilde{K}(iQ)$ independent of the
renormalization point $\kappa$. Finite masses give corrections of
order $m^2(Q)/Q^2$, with $m^2(Q)$ that runs logarithmically and
hence can be ignored, if we are only interested in the lowest moment
of the subtracted SF.

At this point we have expressed the left-hand side of
Eq.~(\ref{ddisp}) as an asymptotic expansion of the form:
\begin{equation}
\Delta\tilde{K}(iQ)\sim
\sum_{n,\nu = 0}^{\infty}
\frac{c_n^{(\nu)}(\kappa)\Delta\langle [O_n]_\kappa\rangle}{Q^{d_n}}
[g^2(Q)]^{\nu+\eta_n}\, ,
\label{asyope}
\end{equation}
where $\Delta\langle[O_n]_\kappa\rangle$ denotes the difference
between the expectation values of $[O_n]_\kappa$ in the ensembles
specified by $(T,\mu)$ and $(T',\mu')$, while the exponent $\eta_n$
and the $Q^2$-independent coefficients $c_n^{(\nu)}(\kappa)$ are
known perturbatively.

We proceed by making an analogous asymptotic expansion of
$\Delta\rho(u)$:
\begin{equation}
\Delta\rho(u)\sim
 \sum_{n=0}^\infty
\frac{[g^2(u)]^{\xi_n}}{u^{2(n+1)}}\,
\sum_{\nu=0}^\infty
a_n^{(\nu)}[g^2(u)]^{\nu}\, .
\label{rholn}
\end{equation}
For notational clarity, we have ignored exponentially suppressed terms
and the fact that there can be more than one $\eta_n$ and $\xi_n$ for
each $n$. We then obtain the sum rules by imposing that the asymptotic
expansion of the right-hand side of Eq.~(\ref{ddisp}), which we get by
inserting Eq.~(\ref{rholn}) in the dispersion integral, matches the
left-hand side obtained by the OPE.
We wish to point out that terms like $g^2T/\sqrt{u^2}$, even
though allowed by dimensional reasons, cannot appear in Eq.~(\ref{rholn}),
because they would generate terms like $g^2T/\sqrt{Q^2}$ in the OPE
series. On the contrary, these kind of terms can be
generated by infrared instabilities in quantities that do not possess
scale separation of the OPE type, e.g. thermodynamical quantities.
Perturbatively, terms like $\ln(u/T)$ would appear in the SF at high-$T$
limit with $u$ fixed, but disappear in the large-$u$ limit with $T$ fixed.
To carry out the dispersion integral, the integration region is splitted
into three intervals, $(0,\lambda^2)$, $(\lambda^2,Q^2)$ and $(Q^2,\infty)$
with $\lambda^2$ large, but otherwise arbitrary. The integration
in the first interval can be carried out naively, while standard
integration by parts techniques have been applied to the last two intervals.
In a long paper~\cite{longpaper} we
present those technical details that make the procedure sketched above
rigorous. In particular, we show how to make the asymptotic expansion of the
dispersion integral (e.g. how Eq.~(\ref{leadingrh}) is obtained from
Eq.~(\ref{rholn})), and how to match it with the asymptotic OPE series of
the correlator.
At the moment we only need to consider the leading terms
in the expansions of, respectively, the left-hand side and the right-hand
side of Eq.~(\ref{ddisp}):
\begin{equation}
[g^2(Q)]^{\eta_n}
\frac{\Delta\langle [O_n]_\kappa\rangle}{Q^{d_n}}
[c_n^{(0)}(\kappa) + c_n^{(1)}(\kappa)g^2(Q)]\, ,
\label{leadinglh}
\end{equation}
\begin{equation}
{1\over Q^2}\biggl[\overline{\Delta\rho}\,
+{a_{0}^{(0)}\over(1-\xi_n)}[g^2(Q)]^{\xi_n-1}\biggr]\, .
\label{leadingrh}
\end{equation}
In Eq.~(\ref{leadingrh}), if $\xi_n>1$, $\overline{\Delta\rho}$
can be shown to be equal to the zeroth moment of the subtracted SF;
note that the zeroth moment of a function whose
asymptotic expansion is Eq.~(\ref{rholn}) is infinite, if $\xi_n\leq 1$,
since $\int_A^\infty \! dx\, x^{-1}(\log x)^{-\xi_n} = \infty$.

For the sake of concreteness, let us examine the consequences of
matching Eq.~(\ref{leadinglh})  and Eq.~(\ref{leadingrh})
in the case we are interested in, {\em i.e.} $n=1$ and
$d_1=2$.

\noindent
If the OPE calculation produces $\eta_n=0$, then $\xi_n$ must be an
integer greater than one, the zeroth moment exists and is given by
\begin{equation}
\int_0^\infty \!\!\! du^2\, \Delta\rho(u) =
c_1^{(0)}(\kappa)\Delta\langle [O_1]_\kappa\rangle\, .
\label{cons}
\end{equation}
If $\eta_n>0$,  then $\xi_n=1+\eta_n$, the zeroth moment is
again finite
and equal to zero:
\begin{equation}
\int_0^\infty \!\!\! du^2\, \Delta\rho(u) = 0\, .
\label{nonc}
\end{equation}

Our main results, Eq.~(\ref{cons}) and Eq.~(\ref{nonc}), can be
expressed in physical terms as follows. The zeroth moment of a SF for
a current $J$ whose OPE expansion yields $\eta_n > 0$ is independent
of $T$ and $\mu$, while the same moment for a current with $\eta_n=0$
changes with $T$ and $\mu$ proportionally to the corresponding
change(s) of the condensate(s) of leading  operator(s). Furthermore,
the zeroth moment clearly is independent of the renormalization point.

Since $\xi_n<1$ or $\xi_n=1$ in Eq.~(\ref{rholn}) produce terms
such as $[g^2(Q)]^{-1}$ or $\ln[g^2(Q)]$ in Eq.~(\ref{leadingrh}),
this possibility, which
implies that the zeroth moment of the SF is infinite, can again be
deduced from the OPE result ($\eta_n < 0$ or the presence of
$\ln[g^2(Q)]$). We remark that, even when the moment is not finite,
the asymptotic expansion is still well defined and,
therefore, the whole approach is still correct~\cite{longpaper}.

At this point several more general comments are
appropriate~\cite{kapusta}:

\noindent (1)
Our derivation relies on the fact that an asymptotically free theory
allows a perturbative expansion at short distances, making practical the
use of the OPE and of the RGE. We understand why only short distance
physics is involved if we realize that the integral over frequencies reduces
Eq.~(\ref{corr}) to the ensemble average of the equal-time commutator
of the currents. Therefore, results such as Eq.~(\ref{cons}) and
Eq.~(\ref{nonc}) are completely determined by the one-loop structure
of the theory and the particular current under exam. Within the
framework of the OPE and of the RGE, the derivation is exact, i.e. there
are not corrections from higher-loop graphs.

\noindent (2)
Flavor, or other non-dynamical quantum numbers, does not change the
expansions at the one-loop level in an essential way and,
therefore, one can derive
analogous sum rules by using other kind of subtractions, instead of
(or in addition to) different $T$ or $\mu$. One such example is given
by the exact Weinberg sum rules at zero $T$~\cite{weinberg} in the
chiral limit.

\noindent (3)
The derivation of sum rules for higher moments of the SF requires the
complete cancelation of all the lower dimensional operator terms with
their complete radiative corrections, not just the leading $g^2(Q)$
corrections; in particular, we also need current
quark mass corrections to the Wilson coefficients. Without appropriate
subtractions, higher moments do not even converge.

\noindent (4)
It is essential to take properly into account the QCD logarithmic
corrections when deriving the exact sum rules, for the
logarithmic corrections not only dictate whether
$\overline{\Delta\rho}$ satisfy Eq.(\ref{cons}) or Eq.(\ref{nonc}),
but also control the very existence of $\overline{\Delta\rho}$.
This procedure is in sharp contrast with the usual QCD sum
rule approach, where the convergence issue is by-passed by applying
the Borel improvement.

\noindent (5)
We believe that the $(T,\mu)$-dependent part of the leading condensate
appearing in Eq.~(\ref{cons}) does not suffer from the infrared
renormalon ambiguity. In fact, only the perturbative term $\tilde{K}_0$
can generate contributions to the leading condensate that are dependent
on the prescription used to regularize these renormalons.
But $\tilde{K}_0$ is independent of $T$ and $\mu$:
any prescription dependence cancels out when we make the subtraction
in Eq.~(\ref{ddisp}). On the contrary, unless we generalize
Eq.~(\ref{ddisp}) and make other subtractions, sum rules that involve
non-leading condensates may be ambiguous.

\noindent (6)
In addition to applying our general results to QCD, we have also
explicitly verified their correctness in a soluble model. A section of
Ref.~\cite{longpaper} is dedicated to illustrate the derivation in
the Gross-Neveu model in the large-$N$ limit,
where we can either calculate the Wilson coefficients, $\beta$-
and $\Gamma$-functions in vector and pseudoscalar channels, or
obtain the exact SF's at arbitrary $(T,\mu)$, and hence directly
verify the sum rules.

  Now let us specialize to QCD and consider four correlation functions:
two involving the non-conserved scalar $J_S=\bar{\psi}\psi$ and
pseudoscalar $J_P=\bar{\psi}\gamma_5\psi$ currents (anomalous dimensions
$\gamma_{J_S}=\gamma_{J_P}=1/4\pi^2$) and two involving the conserved
vector $J_V=\bar{\psi}\gamma_\mu\psi$ and axial-vector
$J_A=\bar{\psi}\gamma_\mu\gamma_5\psi$ currents ($\gamma_V=\gamma_A=0$).
Since the leading operators (dimension four) have nonpositive anomalous
dimensions, the two non-conserved current have
$\eta_n\geq (2\gamma_J-\gamma_n)/2b>0$ and Eq.~(\ref{nonc}) applies,
{\em i.e.} the zeroth moments of their SF's are independent of $T$ and
$\mu$. On the other hand, the two conserved currents have
$\eta_n=0$ and a generalization of Eq.~(\ref{cons})
applies~\cite{longpaper}, where the three dimension-four
operators with zero anomalous dimension  appear.
Two of these operators are Lorentz scalars:
$m\bar{\psi}\psi$ and
$\alpha_s G^2\equiv\alpha_s \text{tr}(G_{\mu\nu}G^{\mu\nu})$,
while the third is the energy-momentum tensor:
\begin{equation}
\theta_{\mu\nu}=
 \frac{1}{4} \bar{\psi}(\gamma_{\mu} i\tensor{D}_{\nu} +
 \gamma_{\nu} i\tensor{D}_{\mu}-m g_{\mu\nu}) \psi
+2\left( \text{tr}(G_{\mu\alpha}G^{\phantom{\nu}\alpha}_{\nu})
-\frac{1}{4} g_{\mu\nu}G^2 \right) \, .
\end{equation}
The sum rule for the vector ($a=6$) and axial-vector ($a=-10$) currents is:
\begin{equation}
\int_0^\infty\!\! du^2\, \Delta\rho(u)=
a\Delta\langle[m\bar{\psi}\psi]\rangle
+{\Delta\langle[\alpha_s G^2]\rangle\over 2\pi}
+8\Delta\langle [\theta_{00}]\rangle\, .
\label{va}
\end{equation}
These exact sum rules should not be affected by instantons, although
the value of the condensates certainly have instanton contributions.
The reason is that the instanton singularities in the Borel-plane
are located on the positive axis starting at $8\pi^2$, and,
therefore, contribute to correlation functions only with
higher order terms in $1/Q^2$~\cite{thooft}.

Finally, let us discuss some of the phenomenological consequences
of these exact sum rules. In the pseudoscalar channel,
$\overline{\Delta\rho}=0$ implies that, in the broken-chiral-symmetry
phase, the change of the pion pole induced by $T$ or $\mu$
is exactly compensated by a corresponding change of the continuum
part of the SF. This result could in principle be verified experimentally,
but more practically at present it implies a strong constraint in
parameterization of SF's. For example, the residue of the pion pole
can not be taken as a parameter independent of the continuum part.
Next let us consider the scalar correlation
function at zero frequency, the chiral susceptibility,
\begin{eqnarray}
\chi(T,\mu)&\equiv&
\int\! d^4x\, \theta(x_0)\langle [J_S(x),J_S(0)]\rangle_{T,\mu}\\
&=&\int_0^\infty\!\! du^2\, {\rho(u;T,\mu)\over u^2}\, ,
\label{chirals}
\end{eqnarray}
which diverges when $(T,\mu)$ approaches the phase boundary,
provided the chiral restoration is a continuous transition.
The divergence of
the chiral susceptibility near phase transition can be only produced
in Eq.(\ref{chirals}) by singularities very close to the origin, when
the exact sum rule $\overline{\Delta\rho}=0$ is simultaneously taken
into account. Thus, when approaching the phase boundary in the
$(T,\mu)$-plane, the threshold of the spectral function vanishes
(since there is no massless poles in the chirally symmetric
phase), and a strong peak develops right above the threshold.
Because in the chirally symmetric
phase the pseudoscalar and scalar channels are degenerate, the same would
also happen to the pseudoscalar SF. This strong peak in the pseudoscalar
and scalar SF's, which is intimately connected with the critical
phenomenon of a diverging susceptibility and correlation length near
the phase transition, can be interpreted as some kind of quasi-particle,
thus confirming the qualitative picture, originally proposed in the
context of the Nambu-Jona-Lasinio model~\cite{kunihiro}, of the
appearance of soft modes near the chiral phase transition.

We believe that similar result hold in the vector and axial-vector
channels, even if our conclusions cannot be as strong as in the previous
case, because of two main differences. The first is that now
$\overline{\Delta\rho}\neq 0$; this problem is not very serious,
since we only need that $\overline{\Delta\rho}$ is not singular
crossing the phase boundary. And this can still be argued by means
of the sum rule in Eq.~(\ref{va}) and the fact that, for a
continuous phase transition, the changes of the thermal energy
$\Delta\langle [\theta_{00}]\rangle$, and of both condensates
$\Delta\langle [m\bar{\psi}\psi]\rangle$ and
$\Delta\langle[\alpha_s G^2]\rangle$ should behave smoothly across
the critical line. The most serious difference is that, in these
channels, we cannot argue on physical grounds that the corresponding
susceptibility diverges. Nonetheless,  there exist lattice
simulations~\cite{gottlieb} showing that the so-called quark number
susceptibility rapidly increases in the transition. Thus it is plausible
that also in the vector channel the SF has a vanishing threshold, and
accumulates strength just above it.

  If the chiral restoration turns out not to be a second order phase
transition (infinite correlation length), but rather a cross-over or
weak first order transition (finite but large correlation), as the
lattice data seem to indicate~\cite{phase}, we expect the same
qualitative features, though less pronounced: a peak develops but
it does not actually diverge at the transition.

  In summary, we used OPE and RGE to derive exact sum rules at
finite $T$ and $\mu$ valid for asymptotically free theories.
We found that, depending on the theory and on the current under
study, the zeroth moment of a SF is either independent of $T$ and
$\mu$, or its change is related to the corresponding changes of
the condensates of operators of lowest dimension. The inclusion
of the logarithmic corrections in our derivation is crucial,
for the logarithmic corrections not only determine whether a
sum rule should obey Eq.(\ref{cons}) or Eq.(\ref{nonc}) but also
control the very convergence of the zeroth moments. These exact
sum rules strongly constrain the shape of SF's in every point of the
$(T,\mu)$-plane, especially near phase transitions.
We urge whoever parameterizes a SF, {\em e.g.}
in QCD sum rule type of calculations or to interpret lattice
simulations, to incorporate these exact constraints. For instance,
the change of the pion residue is not independent of the change of the
continuum part. The present knowledge about the order of the QCD chiral
phase transition together with the information carried by the sum
rule in the scalar channel has important
consequences near the chiral restoration phase
transition. More specifically, we find
that the scalar and pseudoscalar SF's differ drastically from their
zero $(T,\mu)$ shape, and develop a strong peak right above threshold;
we argue that the same might be true for the
vector and axial-vector SF. Therefore, results derived using the same
type of parameterization used at zero $(T,\mu)$ should be taken
with great caution.
In the future, we plan to generalize these results
to baryonic currents and analyse their phenomenological
consequences in greater detail.

This work is supported in part by funds
provided by the U.S.~Department of Energy (D.O.E.) under contracts
\#DE-AC02-76ER03069 and \#DE-FG06-88ER40427.

\end{document}